\documentclass{acm_proc_article-sp}

\usepackage{amssymb}
\usepackage{amsmath}
\usepackage{graphicx}
\usepackage{multirow}
\usepackage{color}
\usepackage{paralist}
\usepackage{subfigure}

\usepackage{listings} 
\usepackage{tabularx}
\usepackage{multirow}
\usepackage[utf8]{inputenc}
\usepackage{mathtools}

\usepackage{algorithm}
\usepackage[noend]{algpseudocode}
\makeatletter
\def\BState{\State\hskip-\ALG@thistlm}
\makeatother

\usepackage{tikz}
\usetikzlibrary{patterns}

\usepackage{ntheorem}
\newtheorem{hyp}{Hypothesis}

\usepackage{pgfplots}

\usepackage{url}

\usepackage{listings}
\usepackage{color} 
\usepackage{enumitem}

\begin{document}

\title{Identifying Web Tables -- \\Supporting a Neglected Type of Content on the Web}

%
%
%
%
%

\numberofauthors{3} 
%
\author{
%
%
\alignauthor
Mikhail Galkin\\
       \affaddr{University of Bonn, Germany}\\
       \affaddr{ITMO University, Saint Petersburg, Russia}\\
       \email{galkin@iai.uni-bonn.de}
\alignauthor
Dmitry Mouromtsev\\
       \affaddr{ITMO University, Saint Petersburg, Russia}\\
       \email{mouromtsev@mail.ifmo.ru}
\alignauthor S\"{o}ren Auer\\
       \affaddr{University of Bonn, Germany}\\
       \email{auer@cs.uni-bonn.de}
}


\maketitle
\begin{abstract}
The abundance of the data in the Internet facilitates the improvement of extraction and processing tools.
The trend in the open data publishing encourages the adoption of structured formats like CSV and RDF. 
However, there is still a plethora of unstructured data on the Web which we assume contain semantics. 
For this reason, we propose an approach to derive semantics from web tables which are still the most 
popular publishing tool on the Web.
The paper also discusses methods and services of unstructured data extraction and processing as well as
machine learning techniques to enhance such a workflow. 
The eventual result is a framework to process, publish and visualize linked open data. 
The software enables tables extraction from various open data sources in the HTML format and an automatic export to the RDF format making the data linked. 
The paper also gives the evaluation of machine learning techniques in conjunction with string similarity functions to be applied in a tables recognition task.\end{abstract}




\section{Introduction}
\label{sec:introduction}
The Web contains various types of content, e.g. text, pictures, video, audio as well as tables. 
Tables are used everywhere in the Web to represent statistical data, sports results, music data and arbitrary lists of parameters.
Recent research \cite{cafarella2008, crestan2011} conducted on the \emph{Common Crawl} census\footnote{Web: \url{http://commoncrawl.org/}} indicated that an average Web page contains at least nine tables. 
In this research about 12 billion tables were extracted from a billion of HTML pages, which demonstrates the popularity of this type of data representation.
Tables are a natural way how people interact with structured data and can provide a comprehensive overview of large amounts and complex information. 
The prevailing part of structured information on the Web is stored in tables.
Nevertheless, we argue that table is still a neglected content type regarding processing, extraction and annotation tools.

For example, even though there are billions of tables on the Web search engines are still not able to index them in a way that facilitates data retrieval. 
The annotation and retrieval of pictures, video and audio data is meanwhile well supported, whereas on of the most widespread content types is still not sufficiently supported. 
Assumption that an average table contains on average 50 facts it is possible to extract more than 600 billion facts taking into account only the 12 billion sample tables found in the Common Crawl. 
This is already \textit{six} times more than the whole \emph{Linked Open Data Cloud}\footnote{Web: \url{http://stats.lod2.eu/}}.
Moreover, despite a shift towards semantic annotation (e.g. via RDFa) there will always be plain tables abundantly available on the Web.
With this paper we want turn a spotlight on the importance of tables processing and knowledge extraction from tables on the Web. 


The problem of deriving knowledge from tables embedded in an HTML page is a challenging research task. 
In order to enable machines to understand the meaning of data in a table we have to solve certain problems:  
\begin{enumerate}[nosep]
\item Search for relevant Web pages to be processed;
\item Extraction of the information to work with;
\item Determining relevance of the table;
\item Revealing the structure of the found information;
\item Identification of the data range of the table;
\item Mapping the extracted results to existing vocabularies and ontologies.
\end{enumerate}

\begin{figure}[!ht]
\centering
\subfigure{
\raisebox{.75\height}{
    \begin{tabular}[!htbp]{|c|c|c|c|}
    \hline
    & P1 & P2 & P3 \\ \hline
   Obj1	&	a1	& B2 & C3 \\ \hline
Obj2 & 	D1	 & E2 & F3 \\ \hline
Obj3	&	G1	& H2 & I3 \\ \hline
Obj4 & 	J1 & K2 & L3 \\ \hline
Obj5	& 	MkS  & NxA & xyz \\ \hline
    \end{tabular}
    \label{fig:subfig1} }
}
\subfigure{
	\includegraphics[width=0.6\textwidth]{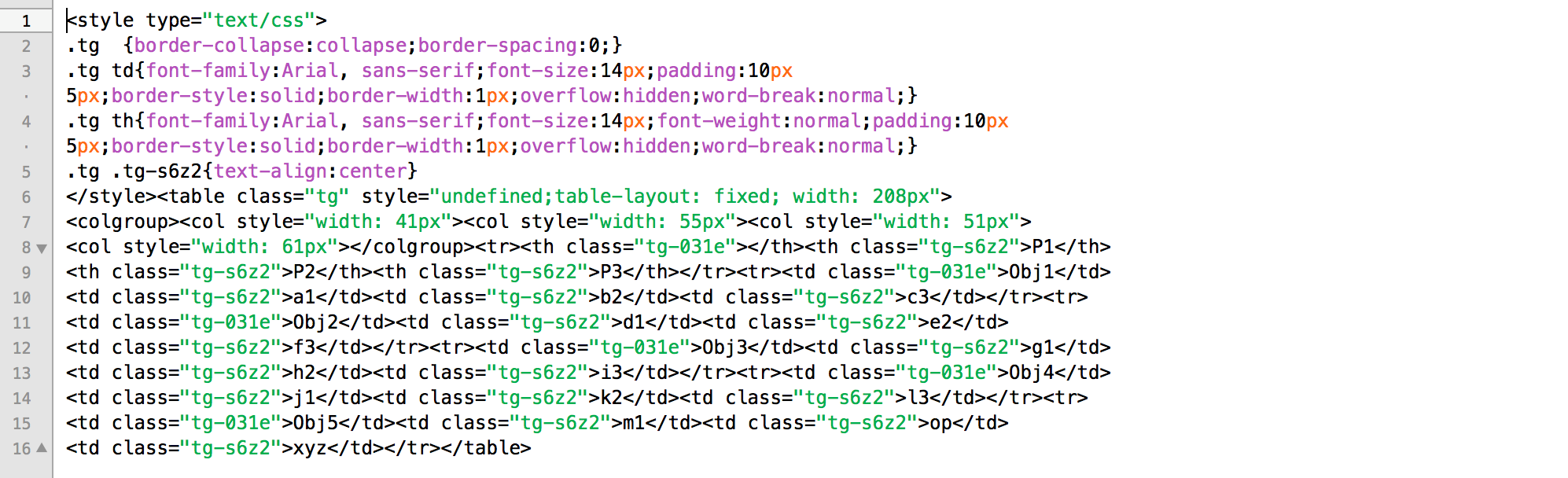}
    \label{fig:subfig2}
}
\caption[Optional caption for list of figures]{Different representations of one table.}
\label{fig:fig1}
\end{figure}

The difference in recognizing a simple table by a human and a machine is depicted in Fig. \ref{fig:fig1}. 
Machine are not easily able to derive formal knowledge about the content of the table.  

The paper describes current methodologies and services to tackle some crucial Web table processing challenges and introduces a new
 approach of table data processing which combines advantages of Semantic Web technologies with robust machine learning algorithms. 
Our approach allows machines to distinguish certain types of tables (genuine tables), recognize their structure (orientation check) and 
dynamically link the content with already known sources.

The paper is structured as follows: 
Section 2 gives an overview of related studies in the field of unstructured data processing.
Section 3 presents Web services which provide the user with table extraction functions.
Section 4 describes the approach and establishes a mathematical ground for a further research.
Section 5 presents used machine learning algorithms and string distance functions. 
Section 6 showcases the evaluation of the approach. 
Finally, we derive conclusions and share plans for future work.

\section{Related Work}
\label{sec:sota}

The Linked Open Data concept raises the question of automatic tables processing as one of the most crucial.
Open Government Data is frequently published in simple HTML tables that are not well structured and lack semantics. 
Thus, the problems discussed in the paper \cite{SPb1} concern methods of acquiring datasets related to roads repair from the government of Saint Petersburg. 
There is also a raw approach \cite{isst2} in information extraction, which is template-based and effective in processing of web sites with unstable markup. 
The crawler was used to create a LOD dataset of CEUR Workshop\footnote{Web: \url{http://ceur-ws.org/}} proceedings. 

A.C. e Silva et al. in their paper \cite{silva2006} suggest and analyze an algorithm of table research that 
consists of five steps: location, segmentation, functional analysis, structural analysis and interpretation of the 
table. 
The authors provide a comprehensive overview of the existing approaches and designed a method for 
extracting data from ASCII tables. 
However, smart tables detection and distinguishing is not considered. 

J. Hu et al. introduced in the paper \cite{hu2002} the methods for table detection and recognition. 
Table detection is based on the idea of edit-distance while table recognition uses random graphs approach. 
M. Hurst takes into consideration ASCII tables \cite{hurst2003} and suggests an approach to derive an abstract 
geometric model of a table from a physical representation. 
A graph of constraints between cells was implemented in order to determine position of cells. 
Nevertheless, the results are rather high which indicates the efficiency of the approach. 
The authors of the papers achieved significant success in structuring a table, but the question of the table content and its semantic is still opened.

D. Embley et al. tried \cite{embley2005} to solve the table processing problem as an extraction problem with an
 introduction of machine learning algorithms. 
However, the test sample was rather small which might have been resulted in overfitting \cite{babyak2004}. 

W. Gatterbauer et al.  in the paper \cite{gatterbauer2007} developed a new approach towards information extraction. 
The emphasis is made on a visual representation of the web table as rendered by a web browser. 
Thus the problem becomes a question of analysis of the visual features, such as 2-D topology and typography. 

Y. A. Tijerino et al. introduced in \cite{tijerino2005} TANGO approach (Table Analysis for Generating Ontologies)
 which is mostly based on WordNet with a special procedure of ontology generation. 
 The whole algorithm implies 4 actions: table recognition, mini-ontology generation, inter-ontology mappings discovery, merging of 
 mini-ontologies. 
 During the table recognition step search in WordNet support the process of table segmentation su that no machine learning algorithms were applied. 

V. Crescenzi et al. introduced in \cite{crescenzi2001} an algorithm RoadRunner for an automatic extraction of 
the HTML tables data. 
The process is based on regular expressions and matches/mismatches handling. 
Eventually a DOM-like tree is built which is more convenient for an information extraction. 
However, there are still cases when regular expressions are of little help to extract any data.

To sum up, there are different approaches to information extraction developed last ten years. 
In our work we introduce an effective extraction and analyzing framework built on top of those methodologies 
combining tables recognition techniques, machine learning algorithms and Semantic Web methods.

\subsection{Existing Data Processing Services}
\label{sec:existing}

Automatic data extraction has always been given a lot of attention from the Web community. 
There are numerous web-services that provide users with sophisticated instruments useful in web scraping, web crawling
and tables processing. 
Some of them are presented below.

\subsubsection{ScraperWiki}
ScraperWiki\footnote{Web: \url{https://scraperwiki.com/}} is a powerful tool based on subscription model that is 
suitable for software engineers and data scientists whose work is connected with processing of large amounts 
of data. 
Being a platform for interaction between business, journalists and developers, ScraperWiki allows 
users to solve extracting and cleaning tasks, helps to visualize acquired information and offers tools to manage
 retrieved data.
Some of the features of the service:
\begin{itemize}[noitemsep,nolistsep]
\item Dataset subscription makes possible the automatic tracking, update and processing of the specified 
dataset in the Internet.
\item A wide range of data processing instruments. For instance, information extraction from PDF documents
\end {itemize}
ScraperWiki allows one to parse web tables in CSV format, but processes all the tables on the page even 
thought they do not contain relevant data, e.g. layout tables. 
Also the service does not provide any Linked Data functionality.

\subsubsection{Scrapy.}
Scrapy\footnote{Web: \url{http://scrapy.org/}} is a fast high-level framework written in Python for web-scraping and 
data extraction. 
Scrapy is spread under BSD license and available on Windows, Linux, MacOS and BSD. 
Merging performance, speed, extensibility and simplicity Scrapy is a popular solution in the industry. 
A lot of services are based on Scrapy, such as ScraperWiki or PriceWiki\footnote{Web: \url{http://www.pricewiki.com/blog/}}. 

\subsubsection{Bitrake.}
Bitrake\footnote{Web: \url{http://www.bitrake.com/}} is a subscription based tool for scraping and processing the data. 
Bitrake offers a special service for those who are not acquainted with programming and provides an API 
for experienced developers written in Ruby and JavaScript. 
One of the distinctive features of the service is a self--developed scraping algorithm with simple filtering and configuration options. 
Bitrake is also used in monitoring and data extraction tasks. 

\subsubsection{Apache Nutch.}
Nutch\footnote{Web: \url{http://nutch.apache.org/}} is an open source web crawler system based on Apache Lucene and written in Java. 
Main advantages of Nutch are performance, flexibility and scalability. 
The system has a modular architecture which allows developers to create custom extensions, e.g. extraction, transforming 
extensions or distribute computing extensions. 
On the other hand, the tool does not have built-in Linked Data functionality, which requires additional development.

\subsubsection{Import.io.}
Import.io\footnote{Web: \url{https://import.io/}} is am emerging data processing service. 
Comprehensive visualization and an opportunity to use the service without programming experience tend Import.io to become 
one of the most wide-spread and user-friendly software. 
The system offers users three methods of extraction arranged by growing complexity: an extractor, a crawler and a connector. 
The feature of automatic table extraction is also implemented but supports only CSV format. 

\section{Concept}
\label{sec:algorithm}
In order to achieve the automatic tables processing certain problems have to be solved:

\begin{enumerate}[nosep]
\item HTML tables search and localization from a URL provided by the user;
\item Computing of appropriate heuristics;
\item Table genuineness check, in other words, check whether a table contains relevant data;
\item Table orientation check (horizontal or vertical);
\item Transformation of the table data to an RDF model.
\end{enumerate}

The importance of correct tables recognition affects the performance of most of web--services. 
The growth of the data on the Web facilitates the data- and knowledge bases updates with such a 
frequency, that does not allow errors, inconsistency or ambiguity. 
With the help of our methodology we aim to address the challenges of automatic \textit{knowledge extraction} and \textit{knowledge replenishment}.

\subsection{Knowledge retrieval} Knowledge extraction enables the creation of knowledge bases and ontologies using the 
content of HTML tables. 
It is also a major step towards five-star open data\footnote{Web: \url{http://5stardata.info/}} 
making the knowledge linked with other datasources and accessible, in addition, in a machine-readable format.
Thus, a correct table processing and an ontology generation is a crucial part of the entire workflow. 
Our framework implements learning algorithms which allow automatic distinguishing between \textit{genuine} 
 and \textit{non-genuine} tables \cite{ukkonen1992}, as well as automatic ontology generation. 
 
We call a table \textit{genuine} when it contains consistent data (e.g. the data the user is looking for) and we 
call a table \textit{non-genuine} when it contains any HTML page layout information or a rather useless 
content, e.g. a list of hyperlinks to other websites within one row or one column.
\\
\subsection{Knowledge acquisition} Knowledge replenishment raises important questions of data updating and deduplication. 
A distinctive feature of our approach is a fully automatic update from the datasource. 
The proposed system implements components of the powerful platform \textit{Information Workbench} 
\cite{haase2011} which introduces the mechanism of \textit{Data Providers}. 
Data Providers observe a datasource specified by a user and all its modifications according to a given schedule. 
Therefore, it enables the replenishment of the same knowledge graph with new entities and facts, which, in 
turn, facilitates data deduplication. 

\section{Formal Definitions}
\label{sec:formal}
The foundation of the formal approach is based on ideas of Ermilov et al. \cite{ermilov2013}
\newdef{definition}{Definition}
\begin{definition}
A table \( \mathcal{T}=(\mathcal{H},\mathcal{N}) \) is tuple consisting of a header \( \mathcal{H}\) and data nodes \( \mathcal{N}\), where:
\begin{itemize}
\item the header \( \mathcal{H} = \{ h_1, h_2, \dots, h_n \} \) is an n-tuple of header elements \( h_i \). We also assume 
that the set of headers might be optional, e.g. \( \exists \mathcal{T}  \equiv  \mathcal{N} \). If the set of 
headers exists, it might be either a row or a column.

\item the data nodes \( \mathcal{N} =  
\begin{pmatrix}
  c_{1,1} & c_{1,2} & \cdots & c_{1,m} \\
  c_{2,1} & c_{2,2} & \cdots & c_{2,m} \\
  \vdots  & \vdots  & \ddots & \vdots  \\
  c_{n,1} & c_{n,2} & \cdots & c_{n,m}
 \end{pmatrix}
\) are a \textit{(n,m)} matrix consisting of \textit{n} rows and \textit{m} columns.
\end{itemize}
\end{definition}


\begin{definition}
The \textit{genuineness} of the table is a parameter which is computed via the function of grid nodes and 
headers, so that \( G_T( \mathcal{N}, \mathcal{H} ) = g_k \in \mathcal{G},  \mathcal{G}=\{ true,flase \} \).
\end{definition}

\begin{definition} 
The \textit{orientation} of the table is a parameter which values are defined in the set \( \mathcal{O}=\{ horizontal, vertical \} \) if and only if \( G_T( \mathcal{N}, \mathcal{H} ) = true\). 
So that \( \exists O_T  \iff G_T \equiv true \) . 
The orientation of the table is computed via a function of grid nodes and headers, so that \( O_T( \mathcal{N}, \mathcal{H} ) = o_k \in \mathcal{O} \). 
If the orientation is horizontal, then the headers are presented as a row, so that \( O_T \equiv horizontal \). 
If the orientation is vertical, then the headers are presented as a column, so that \( O_T \equiv vertical \).
\end{definition}

\begin{table}
\centering
\caption{Horizontal orientation }
\label{tab:horizontal}
\begin{tabular}{|c|c|c|c|c|}
\hline
 H0 & Header 1 & Header 2 &Header 3 & Header 4 \\
 \hline
 Obj 1 &  & & &  \\
 \hline
 Obj 2 & & & &  \\
 \hline
 Obj 3 & & & &  \\
 \hline
 Obj 4 & & & &  \\ \hline
\end{tabular}
\end{table}
\begin{table}
\centering
\caption{Vertical orientation}
\label{tab:vertical}
\begin{tabular}{|c|c|c|c|c|}
\hline
 H0 & Obj 1 & Obj 2 & Obj 3 & Obj 4 \\
 \hline
 Header 1 &  &  & & \\
 \hline
 Header 2 & & & &  \\
 \hline
 Header 3 & & & & \\
 \hline
 Header 4 & & & &\\
 \hline 
\end{tabular}

\end{table}

\begin{definition} A set of \textit{heuristics} \( \mathcal{V} \) is a set of quantitative functions of grid nodes \( \mathcal{N} \) and headers \( \mathcal{H} \) which is used by machine learning algorithms in order to define 
\textit{genuineness} and \textit{orientation} of the given table \( \mathcal{T} \). 

\begin{equation}
   G_T( \mathcal{N}, \mathcal{H}  ) =\begin{cases}
    \textit{true}, &  V_{g1}\in [v_{g1_{min}},v_{g1_{max}}], \dots V_{gm} \\
    \textit{false}, & \text{otherwise}.
  \end{cases}
\end{equation}
where \( V_{g1},...,V_{gm} \in \mathcal{V} \), \( [v_{g1_{min}},v_{g1_{max}}] \) is the range of values of \(V_{g1} \) necessary for the \textit{true} value in conjunction with \(V_{g2},...,V_{gm} \) functions.   

\begin{equation}
   O_T( \mathcal{N}, \mathcal{H}  ) =\begin{cases}
    \textit{horizontal}, &  V_{h1}\in [v_{h1_{min}},v_{h1_{max}}], ... , V_{hn}.\\
    \textit{vertical}, & V_{v1}\in [v_{v1_{min}},v_{v1_{max}}], ... , V_{vl}.
  \end{cases}
\end{equation}
where \( V_{h1},...,V_{hn} \in \mathcal{V} \), \( V_{v1},...,V_{vl} \in \mathcal{V} \), \( [v_{h1_{min}},v_{h1_{max}}] \) is the
 range of values of \(V_{h1} \) necessary for the \textit{horizontal} value in conjunction with \(V_{h2},...,V_{hn} \) 
 functions, \( [v_{v1_{min}},v_{v1_{max}}] \) is the range of values of \(V_{v1} \) necessary for the \textit{vertical} value in
  conjunction with \(V_{v2},...,V_{vl} \) functions.     
\end{definition}

Thus, it becomes obvious, that the \textit{heuristics} are used in order to solve the classification problem 
\cite{mitchell2007} \( X = \mathbb{R}^{n}, Y=\{-1,+1\} \) where the data sample is \( X^l=(x_i,y_i)_{i=1}^l \) and the goal 
is to find the parameters \( w\in \mathbb{R}^{n}, w_0\in \mathbb{R}  \) so that:
\begin{equation}
a(x,w) = sign( \langle x,w \rangle - w_o ).
\end{equation}

We describe heuristics and machine learning algorithms in detail in Section \ref{sec:machine-learning}.
From description logics we define the terminological component \( TBox_T \) as a set of concepts over the set of headers \(\mathcal{H}\).
We define the assertion component \( ABox_T\) as a set of facts over the set of grid nodes \( \mathcal{N} \).
\begin{hyp}
Tables in unstructured formats contain semantics.
\end{hyp}
\begin{equation}
\exists \mathcal{T} | \{  \mathcal{H}_T \iff TBox_T, \mathcal{N}_T \iff ABox_T \}
\end{equation}
In other words, there are tables with the relevant content, which could be efficiently extracted as knowledge. 

The evaluation of the hypothesis is presented in the Section \ref{sec:results}
\subsection{Algorithm Description}

\begin{algorithm}
\caption{The workflow of the system}\label{proc_alg}

\begin{algorithmic}[1]
\Procedure{Workflow}{}
\State $\textit{URI} \gets \text{specified by the user }$
\State $ \text{tables localization}$
\State $ n \gets \text{found tables}$

\BState \( \textit{while } n > 0 \) :
\State $n \gets n-1$.
\If {$ \text{genuine} = true  $}
\State orientation check.
\State RDF transformation.
\EndIf

\State \textbf{goto} \textit{while}.
\EndProcedure

\end{algorithmic}
\end{algorithm}

The valid URL of the website where the data is situated is required from the user. 

The next step is a process of search and localization of HTML tables on a specified website. 
One of the essential points in the process is to handle DOM \( <table> \) leaf nodes in order to avoid nested tables. 
Most of the systems described in \ref{sec:existing} suggest a user with all the extracted tables, whether they are formatting tables or relevant tables. 
In contrast, our approach envisions full automation with the subsequent ontology generation. 

The next step is a computation of heuristics for every extracted table. 
Using a training set and heuristics a machine learning algorithm classifies the object into a genuine or a non-genuine group. 
The input of the machine learning module is a \textit{table trace} -- a unique multi-dimensional vector of computed values of the heuristics of a particular table. 
Using a training set the described classifiers decide, whether the vector satisfies the genuineness class requirements or not. 
If the vector is decided to be genuine the vector then is explored by classifiers again in attempt to define the orientation of the table.

The correct orientation determination is essential for correct transformation of the table data to semantic formats. 
Then it becomes possible to divide data and metadata of a table and construct an ontology. 

If the table is decided to be a non-genuine then a user receives a message where it is stated that a particular table is 
not genuine according to the efficiency of a chosen machine learning method. 
However, the user is allowed to manually mark a table as a genuine which in turn modifies machine learning parameters.

\section{Machine Learning Methods}
\label{sec:machine-learning}
Machine learning algorithms play vital role in the system which workflow is based on appropriate heuristics which in turn 
are based on string similarity functions. 
The nature of the problem implies usage of mechanisms that analyze the content of the table and calculate a set of parameters. 
In our case the most suitable option is implementation of string similarity (or string distance) functions. 

\subsection{String Metrics}
String metric is a metric that measures similarity or dissimilarity between two text strings for approximate string 
matching or comparison. 
In the paper three string distance functions are used and compared.

\textbf{Levenshtein distance} is calculated as a minimum number of edit operations (insertions, substitutions, deletions) 
required to transform one string into another. 
Characters matches are not counted. 

\textbf{Jaro-Winkler distance} is a string similarity metric and improved version of the Jaro distance. 
It is widely used to search for duplicates in a text or a database. 
The numerical value of the distance lies between 0 and 1 which means two strings are more similar the closer the value to 1. 

Being popular in the industry \cite{wang2002} \textbf{n-gram} is a sequence of n items gathered from a text or a speech. 
In the paper n-grams are substrings of a particular string with the length \textit{n}. 
\subsection{Improvements}
Due to the mechanism of tables extraction and analysis certain common practices are improved or omitted. 
Hence particular changes in string similarity functions have been implemented:

\begin{itemize}[noitemsep,nolistsep]
\item The content type of the cell is more important than the content itself. 
Thus it is reasonable to equalize all numbers and count them as the same symbol. 
Nevertheless the order of magnitude of a number is still taken into account. 
For instance, the developed system recognizes 3.9 and 8.2 as the same symbols, but 223.1 and 46.5 would be different with short 
distance between these strings. \\

\item Strings longer than three words have fixed similarity depending on a string distance function in spite of previously
described priority reasons. 
Moreover, tables often contain fields like “description” or “details” that might contain a lot of text which eventually might make a mistake during the heuristics calculations. 
So that the appropriated changes have been implemented into algorithms.
\end{itemize}

\subsection{Heuristics}
Relying on the theory above it is now possible to construct a set of primary heuristics. 
The example table the heuristics mechanisms are explained with is:

\begin{table}[h]
\centering
\caption{An example table}
\begin{tabular}{ |  c  |  c  |  c  |  c  |}
\hline
Name & City & Phone & e-mail  \\
 \hline
 Ivanov I. I. &  Berlin & 1112233 & ivanov@mail.de  \\
 \hline
 Petrov P.P & Berlin & 2223344 & petrov@mail.de  \\
 \hline
 Sidorov S. S. & Moscow & 3334455 & sidorov@ya.ru \\
 \hline
 Pupkin V.V. & Moscow & 4445566 & pupkinv@gmail.com \\
 \hline
\end{tabular}
\end{table}

\subsubsection {Maximum horizontal cell similarity.}
The attribute indicates the maximum similarity of a particular pair of cells normalized to all possible pairs in the row 
found within the whole table under the assumption of horizontal orientation of a table. 
It means the first row of a table is not taken into account because of a header of a table (see Table \ref{tab:horizontal}). 
Having in mind the example table the strings “Ivanov I.I.” and “ivanov@mail.de” are more similar to each other than “Ivanov I.I.” and “1112233”.

The parameter is calculated under the certain rule:
\begin{equation}
maxSimHor= max_{i=2,n}\frac{\sum_{j_1=1}^{m} \sum_{j_2=1}^{m}  dist(c_{i,j_1},c_{i,j_2}) }{m^2}
\end{equation}
where \textit{i} -- a row, \textit{n} -- number of rows in a table, \textit{j} -- a column, \textit{m} -- number of columns in a 
table, \textit{dist()} -- a string similarity function, \( c_{i,j} \) -- a cell of a table. 
\subsubsection{Maximum vertical cell similarity.}
The attribute indicates the maximum similarity of a particular pair of cells normalized to all possible pairs in the column 
found within the whole table under the assumption of vertical orientation of a table. 
It means the first column of a table is not taken into account because in most cases it contains a header (see Table \ref{tab:vertical}). 
According to the example table the parameter calculated for this table would be rather high because it contains pairs of cells with the 
same content (for instance “Berlin” and “Berlin”). 

Using the same designations the parameter is calculated:
\begin{equation}
maxSimVert = max_{j=2,m}\frac{\sum_{i_1=1}^{n} \sum_{i_2=1}^{n}  dist(c_{i_1,j},c_{i_2,j}) }{n^2}
\end{equation}
It is obvious that the greater the maximum horizontal similarity the greater a chance that a table has vertical orientation. 
Indeed, if the distance between values in a row is rather low it might mean that those values are instances of a particular attribute. 
The hypothesis is also applicable to the maximum vertical similarity which indicates possible horizontal orientation.

\subsubsection{Average horizontal cell similarity.}
The parameter indicates average similarity of the content of rows within the table under the assumption of horizontal 
orientation of a table. 
Again, the first row is not taken into account. 
The parameter is calculated under the certain rule:
\begin{equation}
avgSimHor = \frac{1}{n} \sum_{i=2}^{n} \frac{\sum_{j_1=1}^{m} \sum_{j_2=1}^{m}  dist(c_{i,j_1},c_{i,j_2}) }{m^2}
\end{equation}
where \textit{i} -- a row, \textit{n} -- number of rows in a table, \textit{j} -- a column, \textit{m} -- number of columns in a table, \textit{dist()} -- a string similarity function, \textit{c}[{\textit{i,j}] -- a cell of a table. 

The main difference between maximum and average parameters is connected with size of a table. 
Average parameters give reliable results during the analysis of large tables whereas maximum parameters are applicable in case of small tables. 
\subsubsection{Average vertical cell similarity.}
The parameter indicates average similarity of the content of columns within the table under the assumption of vertical 
orientation of a table. 
The first column is not taken into account. 
\begin{equation}
avgSimVert = \frac{1}{m} \sum_{j=2}^{m} \frac{\sum_{i_1=1}^{n} \sum_{i_2=1}^{n}  dist(c_{i_1,j},c_{i_2,j}) }{n^2}
\end{equation}
\subsection{Machine Learning Algorithms}

With the formalization established we are now ready to build classifiers which use apparatus of machine learning. 
Four machine learning algorithms are considered in the paper:

\textbf{Naive Bayes} classifier is a simple and popular machine learning algorithm. 
It is based on Bayes’ theorem with naive assumptions regarding independence between parameters presented in a training set. 
However, this ideal configuration rarely occurs in real datasets so that the result always has a statistical error \cite{mitchell2007}. 

\textbf{A decision tree} is a predictive model that is based on tree--like graphs or binary trees \cite{rokach2008}. 
Branches represent a conjunction of features and a leaf represents a particular class. 
Going down the tree we eventually end up with a leaf (a class) with its own unique configuration of the features and values.  
%

\textbf{k-nearest neighbours} is a simple classifier based on a distance between objects \cite{tou1974}. 
If an object might be represented in Euclidean space then there is a number of functions that could measure a distance between 
these objects. 
If the majority of neighbours of the object belongs to one class than the object would be classified into the same class. 

\textbf{Support Vector Machine} is a non--probabilistic binary linear classifier that tries to divide instances of classes 
presented in a training set by a gap as wide as possible. 
In other words, SVM builds separating surfaces between categories, which might be linear or non-linear \cite{vorontsov}.

\section{Implementation}
The proposed approach was implemented in Java as a plugin for the Information Workbench\footnote{Web: \url{http://www.fluidops.com/en/portfolio/information_workbench/}} platform  developed by fluidOps.
The platform provides numerous helpful APIs responsible for the user interaction, RDF data maintenance, ontology generation, knowledge bases adapters and smart data analysis.
The machine learning algorithms are supplied by \emph{WEKA}\footnote{Web: \url{http://www.cs.waikato.ac.nz/ml/weka/}} -- a comprehensive data mining Java framework developed by the University of Waikato. 
The \emph{SimMetrics}\footnote{Web: \url{http://sourceforge.net/projects/simmetrics/}} library by UK Sheffield University provided string similarity functions.
The plugin is available on a public repository\footnote{Web: \url{https://github.com/migalkin/Tables_Provider}} both as a deployable artifact for the Information Workbench and in source codes.

\section{Evaluation}
\label{sec:results}
The main goal of the evaluation is to assess the performance of the proposed methodology in comparison with the existing solutions.
By the end of the section we decide whether the hypothesis made in Section \ref{sec:formal} is demonstrated or not. 
The evaluation consists of two subgoals -- the evaluation of machine learning algorithms with string similarity functions and the evaluation of the system as a whole.

\subsection{Algorithms Evaluation}
A training set of 400 tables taken from the corpus\footnote{WDC – Web Tables. Web: \url{http://webdatacommons.org/webtables/}} 
 as a result of \cite{cafarella2008} was prepared to test suggested heuristics and machine learning methods. 
In addition, the efficiency of algorithms modifications was estimated during the tests. 
Results are presented on the Table \ref{tab:eval} and Fig. \ref{fig:gencheck},\ref{fig:orcheck}. 
In case of the genuineness check \textit{Precision}, \textit{Recall} and \textit{F--Measure} were computed. 

\begin{table*}
\begin{center}
\caption{Evaluation of the genuineness check}
\label{tab:eval}
\begin{tabular}{| c | l | l | c | c | c | }\hline
   \multicolumn{3}{| c |}{ \textbf{Method}} & \textbf{Precision} & \textbf{Recall} & \textbf{F-Measure}\\ \cline{1-6}
  \multirow{6}{*}{Naive Bayes} & \multirow{2}{*}{Levenshtein} & unmodified & 0.925 & 0.62 & 0.745 \\ \cline{3-3}
  & & modified & 0.93 & 0.64 & 0.76 \\ \cline{2-6}
  & \multirow{2}{*}{Jaro-Winkler} & unmodified & 0.939 & 0.613 & 0.742 \\ \cline{3-3}
  & & modified & 0.939 & 0.617 & 0.744 \\ \cline{2-6}
  & \multirow{2}{*}{n-grams} & unmodified & 0.931 & 0.633 & 0.754\\ \cline{3-3}
  & & modified & 0.937 & 0.643 & 0.763 \\ \cline{1-6}
  \multirow{6}{*}{Decision Tree} & \multirow{2}{*}{Levenshtein} & unmodified & 0.928 & 0.65 & 0.765 \\ \cline{3-3}
  & & modified & 0.942 & 0.653 & 0.76 \\ \cline{2-6}
  & \multirow{2}{*}{Jaro-Winkler} & unmodified & 0.945 & 0.637 & 0.761 \\ \cline{3-3}
  & & modified & 0.946 & 0.64 & 0.763 \\ \cline{2-6}
  & \multirow{2}{*}{n-grams} & unmodified & 0.933 & 0.603 & 0.733 \\ \cline{3-3}
  & & modified & 0.945 & 0.637 & 0.76 \\ \cline{1-6}
  \multirow{6}{*}{kNN} & \multirow{2}{*}{Levenshtein} & unmodified & 0.904 & 0.623 & 0.74 \\ \cline{3-3}
  & & modified & 0.943 & 0.667 & 0.78 \\ \cline{2-6}
  & \multirow{2}{*}{Jaro-Winkler} & unmodified & 0.928 & 0.607 & 0.734\\ \cline{3-3}
  & & modified & 0.941 & 0.64 & 0.762 \\ \cline{2-6}
  & \multirow{2}{*}{n-grams} & unmodified & 0.948 & 0.663 & 0.78 \\ \cline{3-3}
  & & modified & 0.949 & 0.677 & 0.79\\ \cline{1-6}
    \multirow{6}{*}{SVM} & \multirow{2}{*}{Levenshtein} & unmodified & 0.922 & 0.597 & 0.725\\ \cline{3-3}
  & & modified & 0.93 & 0.62 & 0.744 \\ \cline{2-6}
  & \multirow{2}{*}{Jaro-Winkler} & unmodified & 0.924 & 0.61 & 0.735 \\ \cline{3-3}
  & & modified & 0.926 & 0.623 & 0.745 \\ \cline{2-6}
  & \multirow{2}{*}{n-grams} & unmodified & 0.922 & 0.627 & 0.746 \\ \cline{3-3}
  & & modified & 0.927 & 0.637 & 0.755 \\ \cline{1-6}

\end{tabular}
\end{center}
\end{table*}

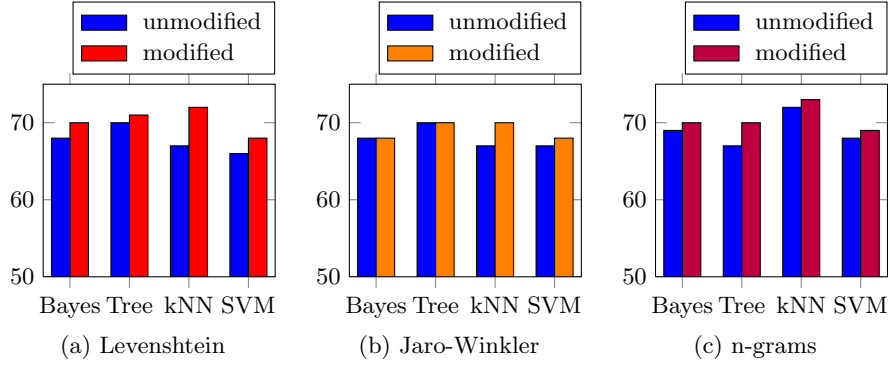
\begin{figure*}
\centering
\subfigure[Levenshtein]{
\begin{tikzpicture}
\begin{axis}[
    ybar = 0pt,
    scale = 0.45,
    /pgf/number format/1000 sep={},
    xticklabels={Bayes, Tree, kNN, SVM},
    xtick = {1,2,3,4,5},
    ymin = 50,
    ymax = 75,
    bar width = 7pt,
    enlarge x limits=0.15,
    legend image code/.code={%
      \draw[#1] (0cm,-0.1cm) rectangle (0.6cm,0.1cm);
    },  
    legend cell align=left,
        legend style={
                at={(1,1.05)},
                anchor=south east,
                column sep=1ex
        }
]
\addplot [fill = blue] coordinates {
    (1, 68) (2, 70) (3, 67) (4, 66) 
};
\addplot [fill=red] coordinates {
    (1, 70) 	(2, 71) (3, 72) (4, 68)
};

\legend{unmodified, modified}
\end{axis}
\end{tikzpicture}
}
\subfigure[Jaro-Winkler]{
\begin{tikzpicture}
\begin{axis}[
    ybar=0pt,
    scale = 0.45,
    /pgf/number format/1000 sep={},
    xticklabels={Bayes, Tree, kNN, SVM},
    xtick = {1,2,3,4,5},
    ymin = 50,
    ymax = 75,
    bar width = 7pt,
    enlarge x limits=0.15,
    legend image code/.code={%
      \draw[#1] (0cm,-0.1cm) rectangle (0.6cm,0.1cm);
    },   
    legend cell align=left,
        legend style={
                at={(1,1.05)},
                anchor=south east,
                column sep=1ex
        }
]
\addplot [fill=blue] coordinates {
    (1, 68) (2, 70) (3, 67) (4, 67) 
};
\addplot [fill=orange] coordinates {
    (1, 68) 	(2, 70) (3, 70) (4, 68)
};
\legend{unmodified, modified}
\end{axis}
\end{tikzpicture}
}
\subfigure[n-grams]{
\begin{tikzpicture}
\begin{axis}[
    ybar=0pt,
   scale = 0.45,
    /pgf/number format/1000 sep={},
    xticklabels={Bayes, Tree, kNN, SVM},
    xtick = {1,2,3,4,5},
    ymin = 50,
    ymax = 75,
    bar width = 7pt,
    enlarge x limits=0.15,
    legend image code/.code={%
      \draw[#1] (0cm,-0.1cm) rectangle (0.6cm,0.1cm);
    },   
    legend cell align=left,
        legend style={
                at={(1,1.05)},
                anchor=south east,
                column sep=1ex
        }
]
\addplot [fill=blue] coordinates {
    (1, 69) (2, 67) (3, 72) (4, 68) 
};
\addplot [fill=purple] coordinates {
    (1, 70) 	(2, 70) (3, 73) (4, 69)
};
\legend{unmodified, modified}
\end{axis}
\end{tikzpicture}
}
\caption{Genuineness check, correctly classified, \%}
\label{fig:gencheck}
\end{figure*}

\begin{figure*}
\centering
\subfigure[Levenshtein]{
\begin{tikzpicture}
\begin{axis}[
    ybar = 0pt,
    scale = 0.42,
    /pgf/number format/1000 sep={},
    xticklabels={Bayes, Tree, kNN, SVM},
    xtick = {1,2,3,4,5},
    ymin = 80,
    ymax = 100,
    bar width = 7pt,
    enlarge x limits=0.125,
    legend image code/.code={%
      \draw[#1] (0cm,-0.1cm) rectangle (0.6cm,0.1cm);
    },   
    legend cell align=left,
        legend style={
                at={(1,1.05)},
                anchor=south east,
                column sep=1ex
        }
]
\addplot [fill=red] coordinates {
    (1, 85) (2, 96) (3, 95.8) (4, 97.6) 
};
\addplot [fill=green] coordinates {
    (1, 85.5) (2, 97) (3, 96.4) (4, 98)
};

\legend{unmodified, modified}
\end{axis}
\end{tikzpicture}
}
\subfigure[Jaro-Winkler]{
\begin{tikzpicture}
\begin{axis}[
    ybar=0pt,
    scale = 0.42,
    /pgf/number format/1000 sep={},
    xticklabels={Bayes, Tree, kNN, SVM},
    xtick = {1,2,3,4,5},
    ymin = 80,
    ymax = 100,
    bar width = 7pt,
    enlarge x limits=0.125,
    legend image code/.code={%
      \draw[#1] (0cm,-0.1cm) rectangle (0.6cm,0.1cm);
    },   
    legend cell align=left,
        legend style={
                at={(1,1.05)},
                anchor=south east,
                column sep=1ex
        }
]
\addplot [fill=red] coordinates {
    (1, 86.2) (2, 95.6) (3, 95.8) (4, 97.6) 
};
\addplot [fill=yellow] coordinates {
    (1, 86.8) (2, 95.4) (3, 95.8) (4, 97.6)
};
\legend{unmodified, modified}
\end{axis}
\end{tikzpicture}
}
\subfigure[n-grams]{
\begin{tikzpicture}
\begin{axis}[
    ybar=0pt,
   scale = 0.42,
    /pgf/number format/1000 sep={},
    xticklabels={Bayes, Tree, kNN, SVM},
    xtick = {1,2,3,4,5},
    ymin = 80,
    ymax = 100,
    bar width = 7pt,
    enlarge x limits=0.125,
    legend image code/.code={%
      \draw[#1] (0cm,-0.1cm) rectangle (0.6cm,0.1cm);
    },   
    legend cell align=left,
        legend style={
                at={(1,1.05)},
                anchor=south east,
                column sep=1ex
        }
]
\addplot [fill=red] coordinates {
    (1, 84.6) (2, 95.8) (3, 95.8) (4, 97.6) 
};
\addplot [fill=cyan] coordinates {
    (1, 85) 	(2, 97) (3, 95.2) (4, 97.6)
};
\legend{unmodified, modified}
\end{axis}
\end{tikzpicture}
}
\caption{Orientation check, correctly classified, \%}
\label{fig:orcheck}
\end{figure*}
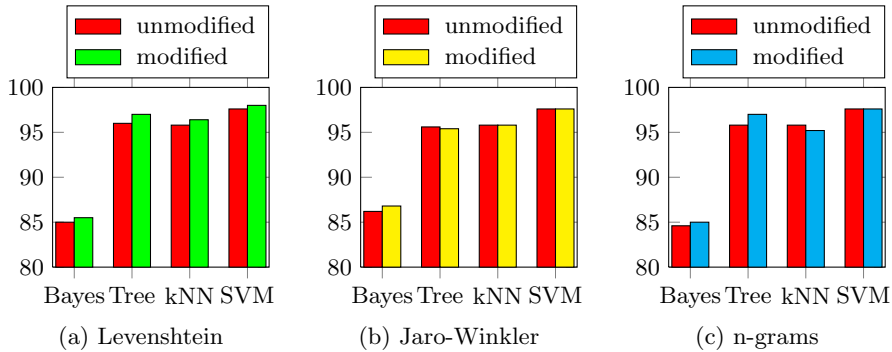
Fig. \ref{fig:gencheck} represents the overall fracture of correctly classified genuine and non-genuine tables w.r.t. used machine learning algorithms and string similarity functions.
The machine learning algorithm based on kNN in conjunction with Levenshtein distance or n-grams demonstrated the 
highest efficiency during the genuineness check. 
A slight increase in efficiency in spite of modifications is observed mostly for kNN. 
It also could be noted that overall results of classification are generally lower in comparison with orientation classification task. 
This may indicate a lack of information about the table structure caused by a small amount of heuristics. 
Development and implementation of more sophisticated numerical parameters is suggested in order to improve the performance of classification.
Hence, the way towards improving overall F--Measure is connected with raising Recall of the approach. 

Fig. \ref{fig:orcheck} indicates the high efficiency of the orientation check task.
Most of the used machine learning methods except Naive Bayes demonstrated close to 100\% results.
 A relatively low result of Naive Bayes regardless of the chosen string similarity function might be explained by a number of assumptions which the method is established on. 
On the one hand the algorithm has the advantage of simplicity and on the other hand it might overlook 
important details which affect the classification process because of such simplicity. 
During the orientation check only genuine tables are considered and assessed. 
Therefore, the eventual result is Precision. 

Having analyzed the efficiency of machine learning methods with string metric mechanisms we decided to apply 
modified kNN in conjunction with Levenshtein distance during the genuineness check process and modified SVM in 
conjunction with Levenshtein distance during the orientation check process.

\subsection{System Evaluation}

The overall performance of the approach is defined as a product of the highest \textit{F--Measure} of the genuineness check and the highest \textit{Precision} of the orientation check, which results in \textit{0.77} or \emph{77\%}.
It is therefore indicating, that we are able to correctly extract knowledge at least from three of given four arbitrary tables.

On Fig. \ref{fig:fig4} the results of tables recognition are presented. 
All the tables that are marked in HTML code of web--pages as tables are coloured in red and blue. 
The tables were extracted from the websites of Associated Press\footnote{Associated Press. Web: \url{http://www.aptn.com/}}, Sports.ru\footnote{Sports.ru. Web: \url{http://www.sports.ru/}} and 
Saint Petersburg Government Portal\footnote{Roads repair dataset | Official Website of Government of Saint 
Petersburg. Web: \url{http://gov.spb.ru/gov/otrasl/tr_infr_kom/tekobjekt/tek_rem/}}. 
According to the theory those tables might be divided in genuine and non-genuine (relevant or irrelevant) groups. 
It might be easily noted that the tables coloured in red use the tag for formatting reasons and do not contain appropriate table data. 
In contrast, the tables coloured in blue are relevant tables which data might be parsed and processed.
\begin{figure}[tb]
        \centering
        \subfigure{\label{fig:a}\includegraphics[width=0.2\textwidth]{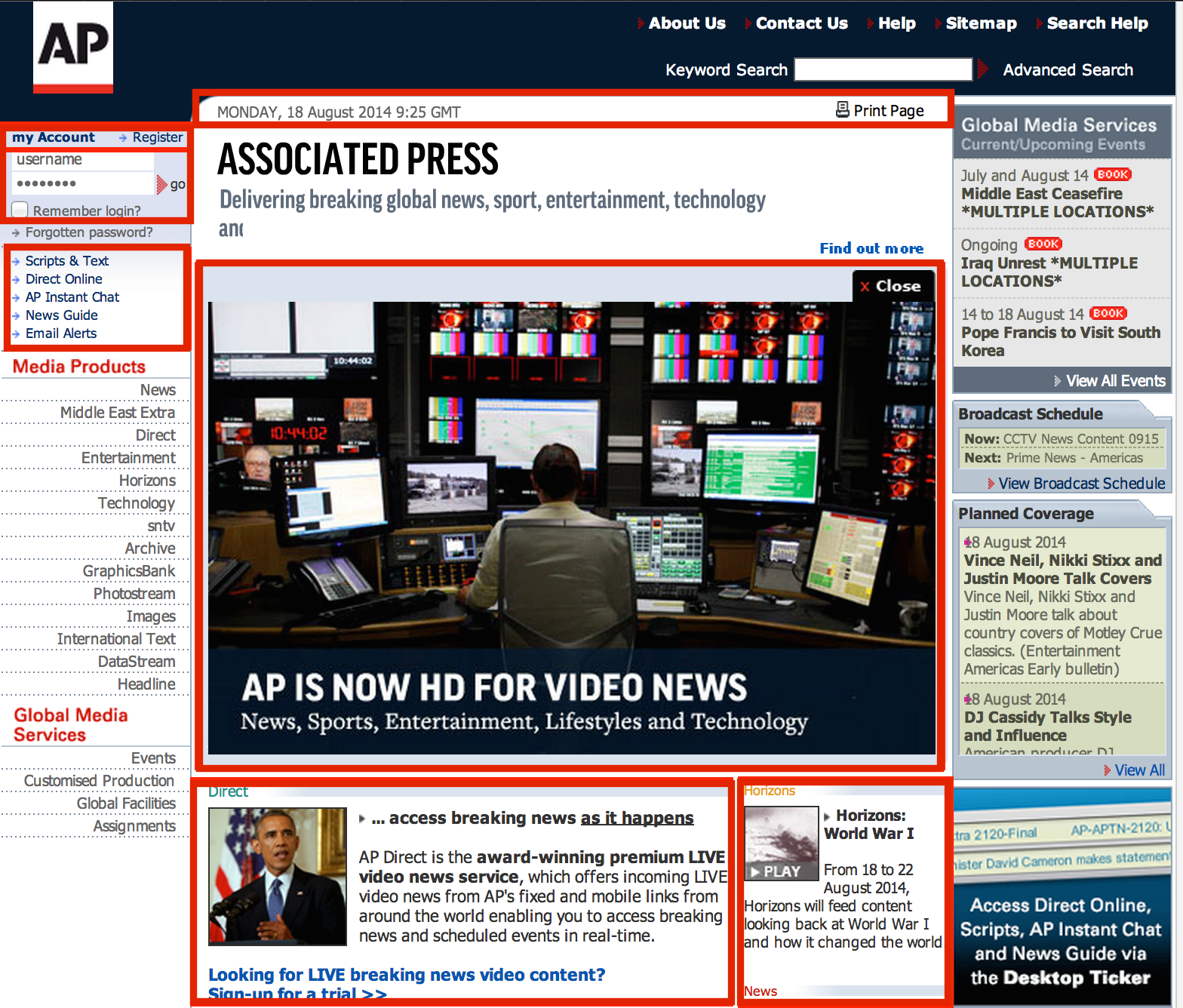}}
        \subfigure{\label{fig:b}\includegraphics[width=0.2\textwidth]{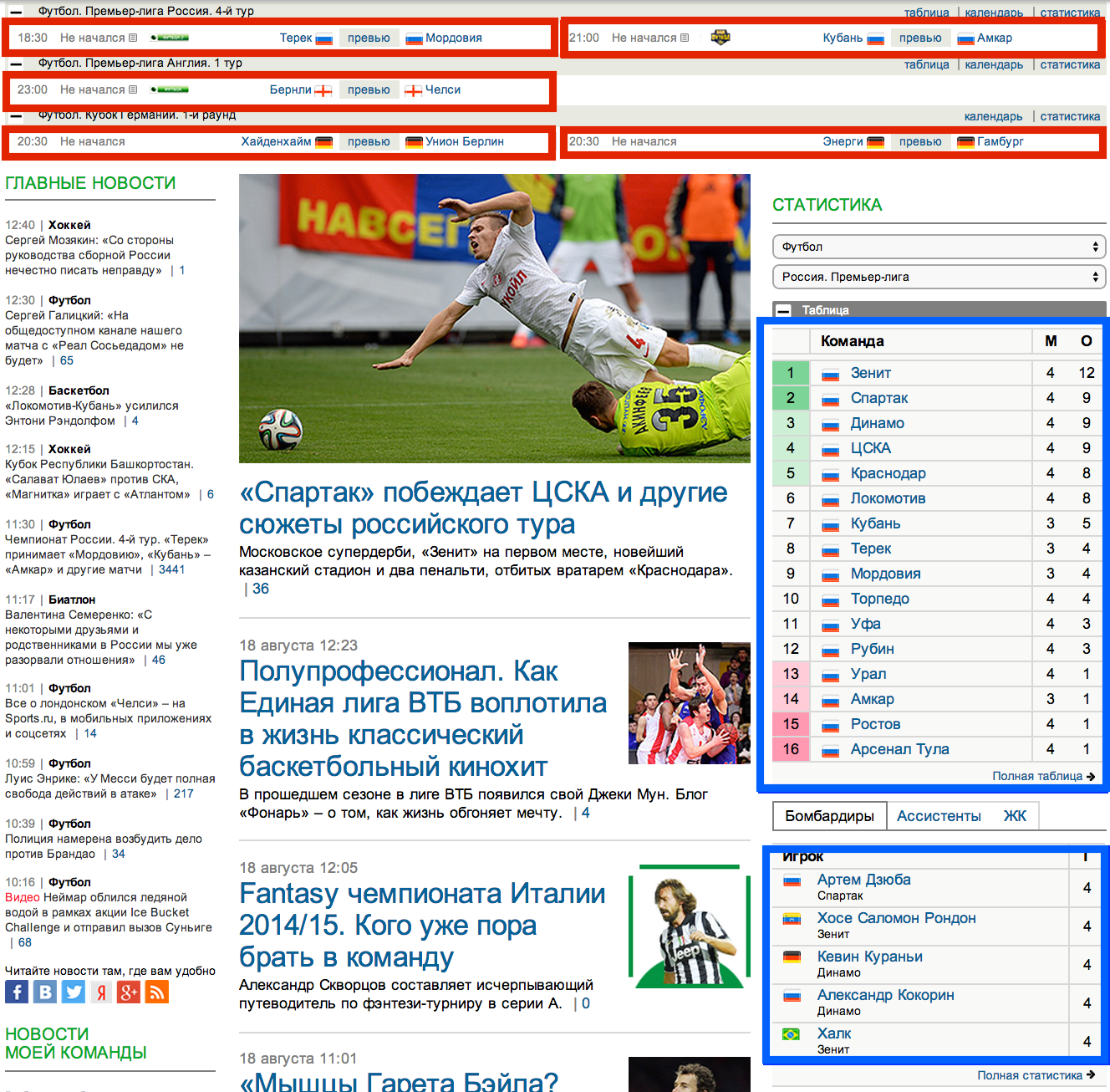}}
        \subfigure{\label{fig:c}\includegraphics[width=0.27\textwidth]{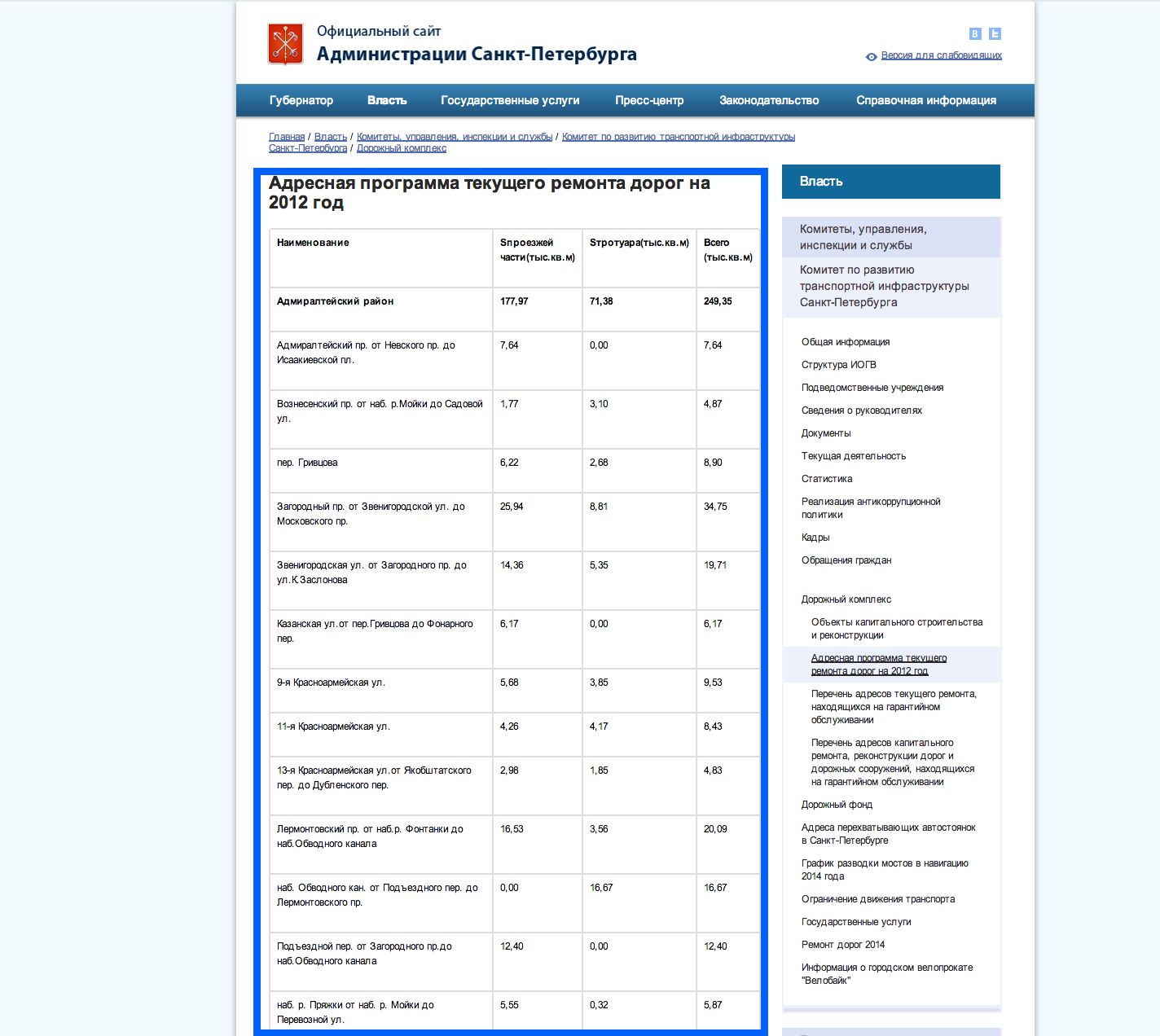}}
        \caption{Table recognition results}
        \label{fig:fig4}
\end{figure}
ScraperWiki was able to extract all the red and blue tables. 
The user therefore should choose relevant tables for a further processing. 
As a counter to ScraperWiki the developed system was able to extract and process only blue genuine tables using appropriate heuristics and machine learning algorithms.

Taking into account the achieved results we consider the hypothesis suggested in Section \ref{sec:formal} demonstrated. 
Indeed, unstructured data contains semantics.
Hence, the next questions are raised.
\textit{How much semantics does unstructured data contain?}
\textit{Is there an opportunity to semantically integrate tables with other types of Web content?}
Answering the questions will facilitate the shift from neglecting the tables towards close integration of all the Web content.

\section{Conclusion and Future Work}
\label{sec:conclusion}

Automatic extraction and processing of unstructured data is a fast--evolving topic in science and industry.  
Suggested machine learning approach is highly effective in table structure analysis tasks and provides the tools for knowledge 
retrieval and acquisition.  

To sum up, the system with the distinctive features was developed:
\begin{enumerate}[nosep]
\item Automatic extraction of HTML tables from the sources specified by a user;
\item Implementation of string metrics and machine learning algorithms to analyze genuineness and structure of a table;
\item Automatic ontology generation and publishing of the extracted dataset;
\item The software takes advantages of Information Workbench API, enabling data visualization, sharing and linking
\end{enumerate}

Future work concerns the question of ontology mapping. 
The datasets to be extracted might be linked with the already existing ones in the knowledge base dynamically during the main workflow, e.g. discovery of the same entities and relations in different datasets. 
It will facilitate the development of the envisioned Web of Data as well as wide implementation of Linked Open Data technologies.


\section{Acknowledgments}
This work was partially financially supported by Government of Russian Federation, Grant 074-U01.

%
\bibliographystyle{abbrv}
\bibliography{lib}  
%
%
\balancecolumns
\end{document}